\documentstyle[prl,aps,epsfig,prbbib]{revtex}
\begin{document}
\twocolumn[\hsize\textwidth\columnwidth\hsize\csname  
@twocolumnfalse\endcsname

\title{Quantitative comparison between theoretical predictions    
and experimental results for the BCS-BEC crossover}
\author{A. Perali, P. Pieri, and G.C. Strinati}
\address{Dipartimento di Fisica, UdR INFM,
Universit\`{a} di Camerino, I-62032 Camerino, Italy}
\maketitle
\date{}

\begin{abstract}
Theoretical predictions for the BCS-BEC crossover of trapped Fermi atoms are 
compared with recent experimental results for the density profiles of $^6$Li.
The calculations rest on a single theoretical approach that
includes pairing fluctuations beyond mean field.
Excellent agreement with experimental results is obtained.
Theoretical predictions for the zero-temperature chemical potential and gap at 
the unitarity limit are also found to compare extremely well with Quantum Monte
Carlo simulations and with recent experimental results.
\vspace{0.3cm}

PACS number(s): 03.75.Hh,03.75.Ss
\end{abstract}
\vspace{0.3cm}
]

\narrowtext

The original theoretical ideas behind the crossover, from the 
Bardeen-Cooper-Schrieffer (BCS) superconductivity with largely overlapping 
Cooper pairs to the Bose-Einstein condensation (BEC) of dilute bosons, date
 back to the pioneering work by Eagles for low-carrier doped superconductors
 \cite{1-Eagles}.
Later seminal papers by Leggett and by Nozi\`{e}res and Schmitt-Rink have
 provided a general framework for the BCS-BEC crossover, both at zero 
temperature in the superfluid phase 
\cite{2-Leggett} and at finite temperature in the normal phase 
\cite{3-NSR}. 
The experimental motivations to these studies came 
from the condensation of excitons in solids \cite{Science-04},
pairing in nuclei \cite{4-nuclei}, and pseudogap in 
high-temperature superconductors \cite{5-HTS}.
No direct quantitative comparison between theory and experiments has, 
however, been possible so far, owing essentially 
to the large number of degrees of freedoms present in these systems. 

Recent experimental advances on the condensation of ultracold trapped Fermi 
atoms \cite{6-Recent-experiments} make it now possible to compare
theoretical predictions for the BCS-BEC crossover 
with experimental results. 
In these systems, the use of a tunable Fano-Feshbach (FF) resonance 
\cite{7-Fano-Feshbach} provides the fermionic attraction that triggers
 pairing, thus enabling one to pass with continuity from weak (BCS) to 
strong (BEC) coupling across the crossover region.
In addition, for broad enough resonances,
the scattering length $a_F$ appears to be the only relevant
 quantity entering the many-body Hamiltonian 
of the interacting Fermi atoms \cite{footholland}.
Experiments on ultracold Fermi atoms thus constitute an ideal testing 
ground for theories which describe the progressive quenching of the fermionic
degrees of freedom into composite bosons.

Comparison of theory with experiments is more interesting for the BCS-BEC
 crossover with Fermi atoms than for the BEC with Bose atoms.
This is because the diluteness condition for Bose gases makes it appropriate
 to describe the bosonic condensate by the Gross-Pitaevskii equation 
\cite{8-GP} and the excitations 
above it by the Bogoliubov approximation \cite{9-Bogoliubov}.
For the BCS-BEC crossover, on the other hand, many-body approximations can 
be controlled only on the two (BCS and BEC) sides of the crossover, where 
the diluteness condition holds for the gas of fermions and composite bosons, 
respectively.
A small parameter to control the many-body approximations is instead lacking 
in the intermediate (crossover) region where the scattering length $a_F$
diverges.

Any sensible theoretical approach to crossover phenomena sets up
a single theory that recovers controlled approximations on both sides of the 
crossover and provides a continuous evolution between them.
It is then of particular relevance that for the BCS-BEC crossover the 
results for the intermediate-coupling regime can be compared with accurate 
results from Quantum Monte Carlo (QMC) simulations \cite{10-QMC} 
performed at the unitarity limit where $a_F$ diverges.
This provides a further stringent test on the validity of a BCS-BEC crossover 
theory.

Characteristic of any BCS-BEC crossover theory is to provide two coupled 
equations for the order parameter $\Delta$ and 
the chemical potential $\mu$ (the latter 
being strongly renormalized from one limit to the other).
In the present theory for the trapped gas, these equations are obtained by 
considering a local-density approximation to the theory of
 Ref.~\onlinecite{11-PPS-04}. The overall chemical potential is replaced 
whenever it occurs by the local quantity 
$\mu({\mathbf r}) = \mu -V({\mathbf r})$ that includes
the trapping potential $V({\mathbf r})$ at position ${\mathbf r}$, 
as discussed in Ref.\onlinecite{ourprl}.
In Ref.~\onlinecite{11-PPS-04}, the theory developed by Popov for a 
weakly-interacting (dilute) superfluid Fermi gas \cite{12-Popov} was 
extended as to include the effects of the collective Bogoliubov-Anderson 
mode \cite{13-Schrieffer} in the diagonal fermionic self-energy.
This was obtained by considering the self-energy (with Nambu notation)
\begin{eqnarray}
\Sigma_{11}(k)&=& - \int\frac{d {\bf q}}{(2\pi)^{3}}\; T 
\sum_{\nu}\Gamma_{11}(q){\mathcal G}_{11}(q-k)\label{sigma-11}\\
\Sigma_{12}(k)&=& - \Delta,
\label{sigma-12}
\end{eqnarray}
where
$\Gamma_{11}(q) = \chi_{11}(-q)/[\chi_{11}(q) \chi_{11}(-q) -
\chi_{12}(q)^{2}]$ is the normal pair propagator with
\begin{eqnarray}
- \, \chi_{11}(q)&=&  \frac{m}{4\pi a_F} + \int \frac{d {\mathbf k}}{(2\pi)^{3}} \, \left[
\; T \sum_{n} \,
{\mathcal G}_{11}(k+q) \, {\mathcal G}_{11}(-k)\right.\nonumber \\ 
& &\phantom{\frac{m}{4\pi a_F}} - 
\left.\frac{m}{{\bf k}^2}\right]\label{A-definition}\\ 
\chi_{12}(q) &=&  \int \! \frac{d {\mathbf k}}{(2\pi)^{3}} \,
\; T \sum_{n} \,
{\mathcal G}_{12}(k+q) \,{\mathcal G}_{21}(-k).
\label{B-definition}
\end{eqnarray}
In these expressions, $q=({\mathbf q},\Omega_{\nu})$ and 
$k=({\mathbf k},\omega_{n})$ (${\mathbf q}$ and 
${\mathbf k}$ are wave vectors,
$\Omega_{\nu}$ and $\omega_n$ bosonic and fermionic 
Matsubara frequencies, respectively), $m$ is the fermion mass, $T$ the
temperature, and ${\mathcal G}_{ij}$  $(i,j=1,2)$ are BCS
Green's functions. (We set $\hbar$=$k_B$=1 throughout.)

In the BCS limit, the quantity $k_F a_F$ ($a_F < 0$) identifies the small
 parameter needed to control the approximations.
Here, the Fermi wave vector $k_F$ is obtained by setting 
$k_F^{2}/(2m)$ equal to the Fermi energy $E_F$ of the non-interacting 
Fermi gas.
The scattering length $a_F$ is varied in experiments 
with trapped Fermi atoms from negative to
 positive values across the FF resonance where $a_F$ diverges.
It was predicted \cite{ourprl,15-PRA-Rapid-03} 
that the crossover from fermionic to bosonic behavior in a trap occurs, in 
practice, within the narrow range $-1\lesssim 
(k_F a_F)^{-1} \lesssim 1$. By our generalization of the 
Popov fermionic approximation, in the strong-coupling limit of the 
fermionic attraction the same fermionic theory is able to describe a 
dilute system of composite bosons within the Bogoliubov approximation
 \cite{9-Bogoliubov}.
In this limit, the small parameter $k_F a_F$ ($a_F > 0$) corresponds to 
the ``gas parameter'' of the dilute Bose gas.

Experiments with trapped Fermi atoms are usually made using anisotropic 
harmonic trapping potentials, having different frequencies
 $\omega_x$, $\omega_y$, and $\omega_z$ for the three axis.
In this case, the Fermi energy $E_F$ equals $(3 N \omega_x \omega_y 
\omega_z)^{1/3}$ where $N$ is the total number of Fermi atoms in the trap.
Within a local-density approximation, the anisotropic problem can be readily
 mapped onto a corresponding isotropic problem
with the same value of the Fermi energy.
Detailed experimental axial density profiles for $^6$Li Fermi atoms were 
reported in Ref.~\onlinecite{14-GRIMM} across the intermediate (crossover) 
region of interest.
In Ref.~\onlinecite{14-GRIMM}, 
$\omega_x = \omega_y = \omega_r$ and $\omega_z = 
\lambda \omega_r$ with $\lambda \ll 1$.
The experimental values of $\omega_r$, $\omega_z$, $a_F$, and $N$ determine 
the value of $(k_F a_F)^{-1}$, to be used for comparison with 
the theoretical calculations.
In addition, for the experiments reported in Ref.~\onlinecite{14-GRIMM} the 
temperature is estimated to be much smaller than $E_F$.
Correspondingly, the theoretical calculations can be performed at $T=0$.

Figure 1 shows the comparison between our theoretical predictions for the 
axial density profiles in the crossover region and the data reported in
 Fig.~4 of Ref.~\onlinecite{14-GRIMM}, 
for three different values of the magnetic 
field $B$ tuning the FF resonance.
The intermediate value $B = 850 G$ (about) corresponds to the position of the 
resonance at which $a_F$ diverges. 
For the two other values $B = 882 G$ and $B = 809 G$, the values of 
the 
scattering length are estimated to be $a_F = - 1.8 \times 10^4 a_0$ and 
$a_F = 8.5 \times 10^3 a_0$, respectively, where $a_0$ is the Bohr radius. 
These values are extracted from the data reported in Fig.~3(a) of 
Ref.~\onlinecite{14-GRIMM}.
The value of $k_F$ is obtained with  
$\omega_r/(2\pi) = 640 Hz$ and $\omega_z/(2\pi) =\sqrt{0.6 \times B[G] + 32} 
Hz$, and depends weakly on 
the estimated value of $N$ which is affected by the largest experimental 
uncertainty (of the order of $50 \%$). 
\vspace{-0.7cm}
\begin{figure}
\centerline{\hspace{1cm}
\epsfig{figure=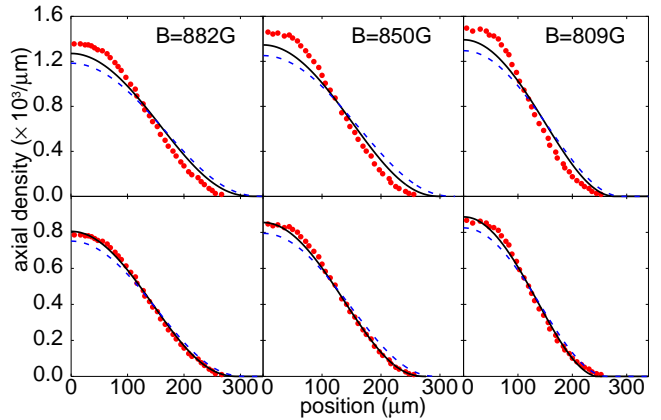,width=10cm,angle=0}\vspace{0.2cm}}
\caption{Comparison between experimental and theoretical axial density 
profiles. Experimental data from Ref. 18
(dots) are shown for three 
diffeent values of the magnetic field $B$ tuning the FF 
resonance. Theoretical results at $T=0$ obtained by our theory 
(full lines) and by BCS mean field (dashed lines) are shown for the 
corresponding couplings $(k_F a_F)^{-1}$ given in the text. 
The upper (lower) panel refers to the estimated number of atoms 
$N = 4 \times 10^5$ ($N = 2.3 \times 10^5$).}
\end{figure}\noindent
This uncertainty affects only the absolute scale of the 
experimental density profiles but not their shape.
The upper panel of Fig.~1 corresponds to the estimated value 
$N = 4 \times 10^5$ given in Ref.~\onlinecite{14-GRIMM}.
With the above values of the frequencies, the coupling $(k_F a_F)^{-1}$ 
is completely determined to be $-0.20$, $0.00$, and $0.43$ from 
left to right, in the order.
Our corresponding theoretical predictions (full lines) compare well with the 
experimental results (dots) in all three cases.
The agreement between theory and experiment becomes almost perfect when 
considering the smaller value $N = 2.3 \times 10^5$ (a value which is 
within the bounds of the experimental uncertainty), as shown in the lower 
panel of Fig.~1.
In this case, $(k_F a_F)^{-1}$ equals $-0.22$, $0.00$, and $0.47$ from 
left to right, in the order \cite{foot820}.

Figure 1 reports also the results of the BCS mean-field calculation 
(dashed lines). 
It is evident that the agreement between theory and experiment is improved 
by our theory (full lines) which includes quantum pairing fluctuations beyond 
mean field.
Nevertheless, the results of the BCS mean field 
appear to be reasonably good in 
comparison with experiment.
This comparison thus verifies a long-standing 
theoretical expectation \cite{2-Leggett} 
that {\em at zero temperature\/} the BCS mean 
field should constitute a reasonable approximation for the whole BCS-BEC 
crossover.

Experimental data over a wider range of $(k_F a_F)^{-1}$ are also available, 
corresponding to an extended variation of the magnetic field about the 
FF resonance.
Figure 2 shows the comparison between our $T=0$ theoretical 
predictions for the normalized root-mean square axial radius (triangles) 
and the data reported in Fig.~3(c) of Ref.~\onlinecite{14-GRIMM} (dots), 
versus the coupling $(k_F a_F)^{-1}$ in the range 
$-1.5 \lesssim (k_F a_F)^{-1}\lesssim 1.5$ 
spanning the whole crossover.
\vspace{-0.5cm}
\begin{figure}
\centerline{\epsfig{figure=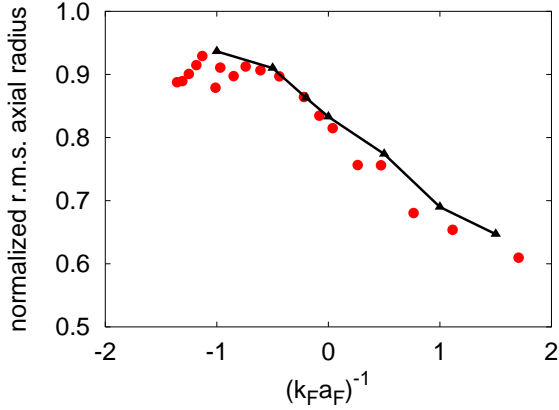,width=8cm,angle=0}\vspace{0.2cm}}
\caption{Comparison between experimental (dots) and theoretical (triangles) 
normalized root-mean square axial radius across the crossover regime. 
Experimental data are taken from Ref. 18. 
The values of $(k_F a_F)^{-1}$ 
and of the non-interacting root-mean square axial radius used also for the 
experimental data are obtained with $N = 2.3 \times 10^5$.}
\end{figure}\noindent
The root-mean square axial radius has been normalized to its
expression $\sqrt{E_F/(4 m \omega_z^2)}$ for non-interacting fermions. 
The values of $(k_F a_F)^{-1}$ and of the non-interacting root-mean square 
axial radius (needed to compare the experimental data with our theoretical 
results) have been obtained with $N = 2.3 \times 10^5$, corresponding to the 
lower panel of Fig.~1. The agreement between theory and experiment is 
remarkably accurate even over this wide coupling range.

As mentioned already, comparison between a crossover theory and the 
experimental data is most compelling in the intermediate-coupling regime, 
due to the lack of a small parameter to control the many-body 
approximations \cite{16-Dyson} 
when the scattering length $a_F$ diverges. 
Special features further occur with a \emph{local\/} theory at the 
unitarity limit $(k_F a_F)^{-1} = 0$.
In particular, this limit corresponds in the isotropic case
to the universal density profile (with $ r=\mid{\mathbf r}\mid$)

\begin{equation}
n(r) = \frac{1}{3 \pi^{2} (1+\beta)^{3/2}} 
[2m (E_F (1+\beta)^{1/2} - V(r))]^{3/2}
\label{density} 
\end{equation}
which depends on the single parameter $\beta$. 
This parameter was introduced experimentally in Ref.~\onlinecite{17-Thomas}, 
and can be extracted theoretically from the ratio 
$\mu/E_F = 1+\beta$ as obtained at the unitarity limit for the homogeneous 
system.
For given values of the harmonic frequencies and of $\beta$, the density 
profile in Eq.~(\ref{density}) thus depends only on the total number of 
atoms $N$ via $E_F$.

Recent $T=0$ QMC simulations \cite{10-QMC} have yielded the 
value $\beta = -0.56 \pm 0.01$.
Our theory gives $\beta = -0.545$, in excellent agreement with these 
simulations.
By contrast, $T=0$ BCS mean field gives $\beta= -0.41$.
Our theoretical value of $\beta$ is also fully consistent with recent 
experimental data, which yield the values $-0.68^{+0.13}_{-0.10}$
 (Ref.~\onlinecite{14-GRIMM}) 
and $-0.55 \pm 0.10$  (Ref.~\onlinecite{18-Salomon}).
The agreement between our theory and QMC simulations is further confirmed 
by comparing the values of the order parameter $\Delta$ at the unitarity limit 
for the homogeneous case.
Our theory gives $\Delta/E_F = 0.53$, while the QMC simulations of 
Ref.~\onlinecite{10-QMC} provide the estimated value $\Delta/E_F = 0.54$.
It is then apparent that our crossover theory, which captures the 
essential physics on the two sides of the crossover, is also able to 
provide quantitatively accurate results in the intermediate-coupling 
regime, where no small parameter can be identified to control the 
many-body problem. 

It is worth commenting that the number of atoms $N = 2.3 \times 10^5$,
utilized by our calculations in the lower panel of Fig.~1 
as well as in Fig.~2, 
was determined by equating the experimental value for the root-mean square 
axial radius at the unitarity limit given in Ref.~\onlinecite{14-GRIMM}, 
to the corresponding value obtained from the universal profile of 
Eq.~(\ref{density}) with $\beta$ taken from the QMC simulations of 
Ref.~\onlinecite{10-QMC}.
We have already seen from Fig.~1 (lower panel) and Fig.~2 that this 
value of $N$ makes the agreement between our theory and the experimental 
data remarkably good. This value of $N$ also improves considerably the 
agreement between the experimental value ($130 nK$) as given in 
Ref.~\onlinecite{14-GRIMM} 
for the bosonic chemical potential $\mu_B$ deep in 
the bosonic regime, and the corresponding theoretical estimate ($125 nK$) 
obtained from the Gross-Pitaevskii theory \cite{8-GP} with the value 
$a_B = 0.6 a_F$ for the dimer-dimer scattering 
length $a_B$ \cite{19-Petrov}.
If one would instead take $N = 4.0 \times 10^5$ (as in the upper 
panel of Fig.~1), a larger value ($155 nK$) would result for the 
theoretical estimate of $\mu_B$. 

In the present approach, the system of fermions maps in strong 
coupling onto a system of composite bosons (dimers), which are described 
by the Bogoliubov theory with a dimer-dimer scattering length $a_B/a_F = 2$.
This value corresponds to the Born approximation for the dimer-dimer 
scattering \cite{20-PS-2000}.
It has been shown \cite{20-PS-2000} that inclusion of higher-order 
dimer-dimer scattering processes within the many-body diagrammatic 
theory decreases this value to $a_B/a_F = 0.75$.
A scattering approach to the four-fermion problem \cite{19-Petrov}
 has given instead the value $a_B/a_F = 0.6$ quoted above.
The difference between the values $0.75$ and $0.6$ originates from additional 
diagrams which were not considered in 
Ref.~\onlinecite{20-PS-2000} albeit they 
survive in the zero-density limit.
It is interesting that the difference between the values $2$ and $0.6-0.75$ 
can be appreciated, by analyzing the experimental results for the 
low-temperature axial density profile reported in Fig.~1(b) of 
Ref.~\onlinecite{14-GRIMM} deep in the 
BEC region (corresponding to $(k_F a_F)^{-1} = 5.44$).
Figure 3 compares the experimental results (dots) directly with the
predictions of the Bogoliubov theory, obtained 
with the alternative values $a_B/a_F = 2$ (dash-dotted line), 
$a_B/a_F = 0.75$ (dashed line), and $a_B/a_F = 0.6$ (full line).
It is evident from this figure that a correct treatment of the dimer-dimer 
scattering improves the comparison with experimental data in this 
extreme BEC limit, where the dimer-dimer 
scattering length is the only relevant interaction parameter.
\vspace{-0.5cm}
\begin{figure}
\centerline{\epsfig{figure=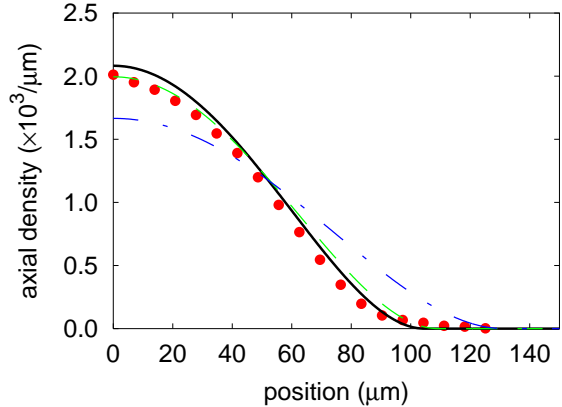,width=8cm,angle=0}\vspace{0.2cm}}
\caption{Comparison between experimental results for the axial density 
profile (dots) and the predictions of Bogoliubov theory,
obtained with $a_B/a_F = 2$ (dash-dotted line), 
$a_B/a_F = 0.75$ (dashed line), and $a_B/a_F = 0.6$ (full line).
Experimental data were obtained in Ref. 18 deep in the BEC 
region, corresponding to $(k_F a_F)^{-1} = 5.44$ when 
$N = 2.3 \times 10^5$.}
\end{figure}\noindent
The analysis of the many-body diagrams made in Ref.~\onlinecite{20-PS-2000} 
suggests, however, that inclusion of diagrams beyond the Born approximation
for the dimer-dimer scattering should become immaterial when approaching the 
crossover region, as these 
diagrams correspond to high-order pairing-fluctuation processes of the 
Fermi system.  

We comment finally that in our approach we have not introduced independent 
bosonic entities besides the fermions. 
The fermions themselves bind, in fact, into composite bosons even at 
finite temperature and density, provided their mutual attraction is 
sufficiently strong.
Correspondingly, the many-body problem contains the two-fermion bound 
state via an effective single-channel problem with scattering length $a_F$,
which can be identified directly with 
the scattering length varied experimentally via the FF resonance.
The excellent agreement with the experimental data shown above demonstrates 
that this single-channel treatment is appropriate to describe the BCS-BEC 
crossover for trapped $^6$Li Fermi atoms.

We are indebted to R. Grimm, J.E. Thomas, 
and K. Levin for discussions, and to R. Grimm for providing us with the 
data files of the figures in Ref.~\onlinecite{14-GRIMM}.
This work was partially supported by the Italian MIUR 
(contract Cofin-2003 ``Complex Systems and Many-Body Problems'').


\vspace{-0.2cm}

\begin{thebibliography}{99}
\vspace{-1.2cm}
\bibitem{1-Eagles} D.M. Eagles, Phys. Rev. {\bf 186}, 456 (1969).

\bibitem{2-Leggett} A.J. Leggett, in \emph{Modern Trends in the Theory of 
Condensed Matter\/}, edited by  A. Pekalski and R. Przystawa, Lecture 
Notes in Physics Vol.115 (Springer-Verlag, Berlin, 1980), p.13.

\bibitem{3-NSR} P. Nozi\`{e}res and S. Schmitt-Rink, J. Low. 
Temp. Phys. {\bf 59}, 195 (1985).

\bibitem{Science-04} C.W. Lai, J. Zoch. A.C. Gossard, and D.S. Chemla,
Science {\bf 303}, 503 (2004), and references quoted therein.

\bibitem{4-nuclei} M. Baldo, U. Lombardo, and P. Schuck, 
Phys. Rev. C {\bf 52}, 975 (1995).

\bibitem{5-HTS} See, e.g., V.M. Loktev, R.M. Quick, and S.G. Sharapov, 
Phys. Rep. {\bf 349}, 1 (2001).

\bibitem{6-Recent-experiments} M. Greiner, C.A. Regal, and D.S. Jin, 
Nature {\bf 426}, 537 (2003);
S. Jochim, M. Bartenstein, A. Altmeyer, G. Hendl, S. Riedl, C. Chin, 
J. Hecker Denschlag, and R. Grimm, Science {\bf 302}, 2101 (2003);
M.W. Zwierlein, C.A. Stan, C.H. Schunck, S.M.F. Raupach, S. Gupta, 
Z. Hadzibabic, and W. Ketterle, Phys. Rev. Lett. {\bf 91}, 250401 (2003);
C.A. Regal, M. Greiner, and D.S. Jin, Phys. Rev. Lett. {\bf 92}, 040403 (2004).

\bibitem{7-Fano-Feshbach} U. Fano, Nuovo Cimento {\bf 12}, 156 (1935);
Phys. Rev. {\bf 124}, 1866 (1961);
H. Feshbach, Ann. Phys. {\bf 19}, 287 (1962);
S. Inouye, M.R. Andrews, J. Stenger, H.-J. Miesner, D.M. Stamper-Kurn, 
and W. Ketterle, Nature {\bf 392}, 151 (1998).

\bibitem{footholland} In contrast, for narrow resonances a bose-fermion
model should be more appropriate. See, {\em e.g.},
J.N. Milstein, S.J.J.M.F. Kokkelmans, and M.J. Holland, 
Phys. Rev. A {\bf 66}, 043604 (2002);
Y. Ohashi and A. Griffin, Phys. Rev. A {\bf 67}, 033603 (2003).

\bibitem{8-GP} F. Dalfovo, S. Giorgini, L.P. Pitaevskii, and S. Stringari, 
Rev. Mod. Phys. {\bf 71}, 463 (1999).

\bibitem{9-Bogoliubov} Cf., e.g., A.L. Fetter and J.D. Walecka, 
\emph{Quantum Theory of Many-Particle Systems\/} (McGraw-Hill, New York, 1971).

\bibitem{10-QMC} J. Carlson, S.-Y. Chang, V.R. Pandharipande, and K.E.
Schmidt, Phys. Rev. Lett. {\bf 91}, 050401 (2003).

\bibitem{11-PPS-04} P. Pieri, L. Pisani, and G.C. Strinati, 
Phys. Rev. Lett. {\bf 92}, 110401 (2004).

\bibitem{ourprl} A. Perali, P. Pieri, L. Pisani, and G.C. Strinati, 
Phys. Rev. Lett. (2004), in press.

\bibitem{12-Popov} V.N. Popov, \emph{Functional Integrals and 
Collective Excitations\/} (Cambridge University Press, Cambridge, 1987). 

\bibitem{13-Schrieffer} J.R. Schrieffer,
 \emph{Theory of Superconductivity\/} (W.A. Benjamin, New York, 1964), 
Chapter 8.

\bibitem{15-PRA-Rapid-03} A. Perali, P. Pieri, and G.C. Strinati, 
Phys. Rev. A {\bf 68}, 031601(R) (2003).

\bibitem{14-GRIMM} M. Bartenstein, A. Altmeyer, S. Riedl, S. Jochim, 
C. Chin, J. Hecker Denschlag, and R. Grimm, Phys. Rev. Lett. {\bf 92}, 
120401 (2004). 

\bibitem{foot820} More recent data \cite{18-Salomon,newKett} place the
position of the resonance at about 820$G$. With these data, the best estimated
value of $N$ is about $1.65\times 10^5$, to which there correspond the values
-1.02, -0.55, and 0.32 for $(k_F a_F)^{-1}$ in Fig.1 (from left to right).
Even with these new values of $N$ and the scattering length, we have
verified that the agreement between the experimental and theoretical 
density profiles is very good.

\bibitem{18-Salomon} T. Bourdel, L. Khaykovich, J. Cubizolles, 
J. Zhang, F. Chevy, M. Teichmann, L. Tarruell, S.J.J.M.F. Kokkelmans, and 
C. Salomon, cond-mat/0403091.

\bibitem{newKett} M.W. Zwierlein, C.A. Stan, C.H. Schunck, S.M.F. Raupach, 
A.J. Kerman, and W. Ketterle, cond-mat/0403049. 

\bibitem{16-Dyson} F. Dyson, Nature {\bf 427}, 297 (2004).

\bibitem{17-Thomas} K.M. O'Hara, S.L. Hemmer, M.E. Gehm, S.R. Granade, 
and J.E. Thomas, Science {\bf 298}, 2179 (2002).

\bibitem{19-Petrov} D. Petrov, C. Salomon, and G. Shlyapnikov, 
cond-mat/0309010.

\bibitem{20-PS-2000} P. Pieri and G.C. Strinati, Phys. Rev. B {\bf 61}, 
15370 (2000), and cond-mat/9811166. 
\vspace{-1.0cm}

\end{thebibliography}
\end{document}